\documentclass[aps,preprint,nofootinbib,preprintnumbers]{revtex4}

\usepackage[dvips]{color}
\usepackage[normalem]{ulem}
\usepackage{amsmath}
\usepackage{enumerate}
\usepackage{amsfonts}
\usepackage{epsfig}

\newcommand{\be}{\begin{equation}}
\newcommand{\ee}{\end{equation}}
\newcommand{\bea}{\begin{eqnarray}}
\newcommand{\eea}{\end{eqnarray}}

\newcommand{\bln}{\begin{align}}
\newcommand{\eln}{\end{align}}
\newcommand{\bst}{\begin{split}}
\newcommand{\est}{\end{split}}
\newcommand{\bi}{\begin{itemize}}
\newcommand{\ei}{\end{itemize}}
\newcommand{\ben}{\begin{enumerate}}
\newcommand{\een}{\end{enumerate}}

\def\ov{\over}
\def\le{\left}
\def\ri{\right}
\def\ha{{1\over 2}}
\def\lam{{\lambda}}
\def\Lam{{\Lambda}}

\def\al{{\alpha}}

\def\sgn{{\rm sgn}}
\def\det{{\rm det}}

\def\NN{{\cal N}}
\def\th{{\theta}}

\def\Om{{\Omega}}
\def \th{{\theta}}

\def \lam {\lambda}

\def\sig{{\sigma}}
\def\ep{{\epsilon}}
\def\apr{{\alpha'}}
\newcommand{\p}{\partial}

\newcommand\Ga{{\ensuremath{{\Gamma}}}}

\def\lam{{\lambda}}

\def\eeq{\end{equation}}

\newcommand\sI{{\ensuremath{{\mathcal I}}}}

\newcommand\sL{{\ensuremath{{\mathcal L}}}}

\newcommand\sN{{\ensuremath{{\mathcal N}}}}
\newcommand\sO{{\ensuremath{{\mathcal O}}}}

\newcommand{\RR}{\mathbb{R}}

\newcommand{\yI}{y_\mathcal{I}}
\newcommand{\yO}{y_\mathcal{O}}

\newcommand{\eps}{\epsilon}

\begin{document}

\title{
Condensed matter physics of a strongly 
coupled gauge theory with quarks:
some novel features of the phase diagram}

\preprint{CTP-LNS 4007}

\author{Thomas Faulkner and Hong Liu}

\affiliation{Center for Theoretical Physics, \\
Massachusetts
Institute of Technology, \\
Cambridge, MA 02139
}


\bigskip
\bigskip

\vspace{1cm}

\begin{abstract}
We revisit the phase diagram of the $\NN=4$ $SU(N_c)$ super-Yang-Mills 
theory coupled to
$N_f$  fundamental ``quarks''  at strong coupling using the gauge-gravity
correspondence.
We show that in the plane of temperature v.s. baryon chemical potential 
there is a critical
line of third order phase transition which ends at a tricritical point 
after which the transition
becomes first order. Close to the critical line
there is an intriguing logarithmic behavior, which cannot 
follow from a 
mean field type of analysis.
We argue that on the string theory side the third order phase transition 
is driven by the condensation of
worldsheet instantons and that this transition 
might become a smooth crossover at finite 't 
Hooft coupling.

\end{abstract}

\maketitle
\newpage
\section{Introduction}
\label{intro}

Fascinating new dynamical phenomena can appear when the coupling(s)
of a system becomes strong. Familiar examples
include color confinement and dynamical chiral symmetry breaking in QCD and
high $T_c$ superconductivity in various condensed matter systems.
Yet strongly coupled systems are hard to solve and intuitions gained from weakly coupled systems normally do not apply.

The AdS/CFT correspondence~\cite{AdS/CFT}
has provided important tools for studying many strongly coupled systems
by relating them to classical gravity (or string) systems. A remarkable feature of the duality is that highly
dynamical, strongly coupled phenomena in gauge theories
can often be understood on the gravity side using simple intuitive geometric pictures.
For example, confinement is reflected in the fact that a fundamental string, which represents an unbounded quark in the boundary gauge theory, has no place to end in the bulk geometry and thus cannot exist. 
New strongly coupled phenomena may be waiting to be discovered,
and AdS/CFT maybe just the tool needed to discover it.

In this paper we are interested in understanding the phase structure of a large $N_c$ gauge theory coupled to a small number $N_f$ fundamental quarks at strong coupling from gravity. More precisely, we consider
$\sN=4$ SYM theory with a gauge group $SU(N_c)$ coupled to $N_f$ ($\sN=2$) hypermultiplets in the fundamental representation of $SU(N_c)$, which in the limit of $N_c \gg N_f$ and strong 't Hooft coupling can be described in terms of $N_f$ probe D7-branes in the AdS$_5 \times S_5$ geometry~\cite{Karch:2002sh}\footnote{See~\cite{Erdmenger:2007cm} for a review of the system.}. Comparing to QCD, the system has the following distinct features:

\ben

\item $N_c \gg N_f$, while in QCD $N_c \sim N_f$.

\item There are both fermionic and {\it bosonic} ``quarks'', which are charged under a $U(1)_B$
baryon number symmetry.

\item The beta-function for gauge coupling is zero to leading order in $N_f/N_c$.
The scale of the system is set by the quark mass $m_q$.

\een
Despite many important differences from QCD, the system appears to be rather interesting in its own right and provides a nice laboratory for studying strongly coupled quark-gluon systems under extreme conditions.
At zero baryon density and finite temperature
it has been used to model heavy quark mesons
in QCD \cite{Dusling:2008tg,Ejaz:2007hg}. At finite baryon density and zero
temperatures some novel dynamical features 
were recently found in \cite{Karch:2008fa}.

\begin{figure}[h!]
\centerline{\hbox{\psfig{figure=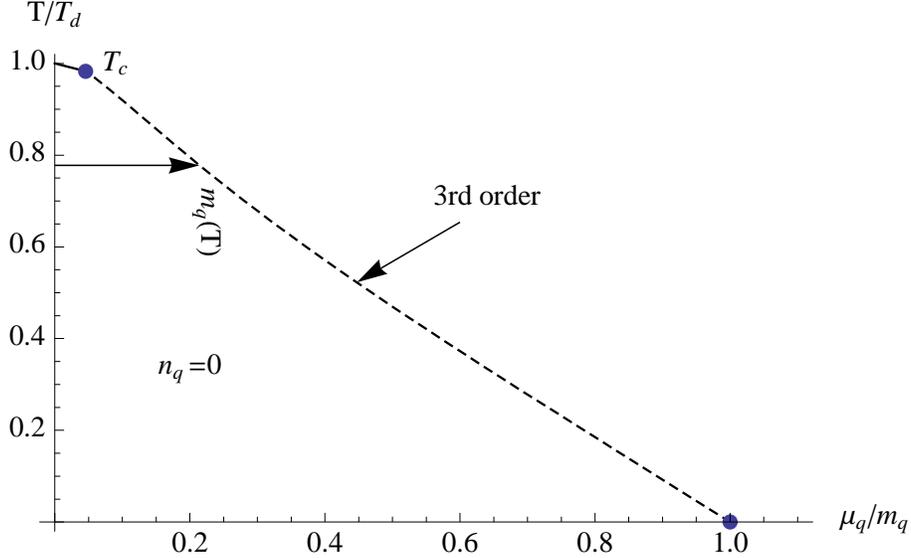,width=12cm
}}}
\caption{The phase diagram in the $\mu_q-T$ plane. The dashed
line indicates a continuous transition. The transition
line lies exactly on the curve $m_q^{(T)}$ until a critical
temperature $T_c$ very close (but not equal) to the
dissociation temperature $T_d$.
\label{fig:phase}}
\end{figure}

The phase diagram of the system at  finite temperature $T$ and baryon chemical potential\footnote{In this paper we will use the terms baryon charge density and quark charge density interchangeably without a factor $1/N_c$ between them.
In other words we take the $U(1)_B$ charge of a quark (anti-quark)
to be 1(-1).
}  $\mu_q$ can be worked out by studying possible configurations of D7-branes in the background geometry (which is a black hole in AdS$_5 \times S_5$)
and has been studied by various authors in~\cite{Mateos:2006nu,Nakamura:2007nx,Ghoroku:2007re,Karch:2007br,Mateos:2007vc,Babington:2003vm,Kirsch:2004km,
Albash:2006ew}
\footnote{
For related studies of the same system see~\cite{Nakamura:2006xk,Kobayashi:2006sb,Erdmenger:2008yj} and 
for related studies of a confining theory see \cite{Horigome:2006xu}}. The results can be summarized as follows (see Fig.~\ref{fig:phase}):

\ben

\item There is a transition curve in the $\mu_q-T$ plane which intersects with the horizontal axis at $\mu_q = m_q$ with $m_q$ the bare quark mass, and with the vertical axis at some temperature $T = T_d$. 
 Except for a small region near the vertical axis, the transition curve is given by
  \be \label{cur}
  \mu_q = m_q^{(T)} (T)
  \ee
  where $m_q^{(T)}$ is the effective quark mass at finite temperature (it decreases with temperature).\footnote{For a precise definition of $m_q^{(T)}$ see discussion around eqs.~\eqref{ybd}--\eqref{quM1} in sec.~\ref{sec:gen}.} In particular, the transition
  is second order~\cite{Karch:2007br} along the horizontal axis ($T=0$) and first order along the vertical axis ($\mu_q =0$)~\cite{Mateos:2006nu}. The transition 
  along the vertical axis at $T=T_d$ has been interpreted as a dissociation transition for mesons~\cite{Mateos:2006nu,Hoyos:2006gb,Mateos:2007vn}.

\item  The region inside the curve in the $\mu_q-T$ plane is described by a D7-brane embedding which lies entirely outside the black hole (Minkowski-type embedding), while the region outside the curve is described by a configuration in which part of the D7-branes falls into the black hole (black hole-type embedding). In terms of the boundary gauge theory, the baryon number density is zero inside the curve and becomes nonzero outside (except along the
vertical axis $\mu_q$ where the baryon density is always zero.)


\een

In this paper we improve on the above description and show that the transition along the curve~\eqref{cur} is a {\it continuous} (3rd order) phase transition which connects to a first order line (near the vertical axis) through a tricritical point whose location we identify precisely, see Fig~.\ref{fig:phase}.

While our conclusions of a continuous phase transition along~\eqref{cur} 
are consistent with the analysis of~\cite{Ghoroku:2007re} who noted
a phase transition along the $m_q^{(T)}$ line at zero density 
they are different 
from a later discussion in~\cite{Mateos:2007vc} where a first order transition was noted.  In~\cite{Mateos:2007vc} the relation between charge density 
and temperature was studied
numerically near the transition line and a very small discontinuity 
in density was noticed, due to a change in dominance between
a Minkowski type embedding with zero density and a black
hole embedding. 
It appears to us that the discontinuity likely has to do with the numerical accuracy of their
calculation. We do not find a discontinuity in charge density along the transition line (see equation~\eqref{criB} below)\footnote{We have obtained our results both analytically and numerically with agreement.}. More importantly, we have identified a clear physical reason for the phase transition, which indicates that it should not be considered as an exchange of dominance between different embedding solutions, which was behind the reasoning of the conclusion in~\cite{Mateos:2007vc}. As observed in~\cite{Faulkner:2008qk}, for Minkowski type embedding at finite temperature, there are worldsheet instanton corrections to the leading order Dirac-Born-Infeld (DBI) action for the D7-branes.
While for a Minkowski type embedding, particle excitations on the D7-branes are not sensitive to the value of the chemical potential, the worldsheet instantons which correspond to semi-classical strings stretched between D7-branes and
the black hole do. In particular, when the baryon chemical potential exceeds the value~\eqref{cur}, the instanton actions exponentially dominate over the DBI action in the large 't Hooft coupling limit and induce an instability.
The consequence of the instability is that instantons ``condense'' and generate
a genuine neck between the D7-branes and the black hole. Thus beyond~\eqref{cur}
the Minkowski-type embedding cannot exist and should be replaced by a black hole type
embedding.\footnote{Note that this reasoning does not by itself imply that the transition
should be continuous, it only indicates that the transition is not an exchange of dominance as is normally the case for a first order transition. Our explicit calculation
shows that there is no jump in the baryon charge density across the transition line
in the infinite 't Hooft coupling limit.} The transition also has a simple interpretation
in the gauge theory. As discussed in~\cite{Faulkner:2008qk}, the instantons can be  interpreted in boundary gauge theory as thermal medium quarks. When $\mu_q$ exceeds the value of~\eqref{cur}, the quarks have negative free energies and will condense and generate a finite charge density, although it is not
clear whether it is fermionic quarks or bosonic quarks which are
condensing.

More explicitly, we find along the critical line~\eqref{cur} (approaching it from above), the baryon charge density and the chemical potential are related by
 \be \label{criB}
 \mu_q - m_q^{(T)} = -B (T) \ep \log \ep + A (T) \ep + O(\ep^2)
 \ee
where $\ep$ is the quark charge density (normalized to be dimensionless) and $A(T), B(T)$
are some functions of temperature. In the zero temperature limit, $B(T)$ goes to zero and $A(T)$ goes to a finite constant, and one recovers the second transition at $T=0$~(found in~\cite{Karch:2007br}), where various exponents are given by their mean field values.
At any finite temperature, however, it is always the logarithmic term on RHS
of~\eqref{criB} that dominates at small enough densities and as a result the transition becomes third order.  The logarithmic behavior does not appear to have a
mean field counterpart, and thus this is an example of continuous phase transition from gravity which does not obey the Landau-Ginsburg behavior.
Such log terms however do appear to be a common feature
of the renormalization group analysis of
condensed matter systems at their upper critical dimension.

There also exists a temperature $T_c$ at which $B(T_c) =0$ and beyond which $B(T) < 0$. At $T_c$ the transition is again second order. For $T > T_c$, the transition becomes first order since for a given $\mu_q$ close to, but {\it smaller} than $m_q^{(T)}$, now there are two black hole type embeddings with $\ep  \neq 0$ in addition to the Minkowski type embedding. Connecting a continuous critical line and a first order transition line, the point $(T=T_c, \mu_q = m_q^{(T)} (T_c))$ is thus a tricritical point.
Again the critical behavior near the tricritical point is {\it not} the same as that of a Landau-Ginsburg type effective theory.

While it is natural to connect the first transition near the tricritical point with that near the vertical axis at $\mu_q =0$, our approximation in~\eqref{criB} which applies to small densities does not extend all the way to $\mu_q =0$. Thus that part of the phase diagram remains a conjecture at the moment.
Also numerical work at \emph{finte} density 
by other authors \cite{Mateos:2007vc} suggest this should be
the case, however a conclusive argument remains to be made.

The phase diagram in Fig.~\ref{fig:phase} 
is for the $\lam = \infty$ limit, where the quark charge
density is identically zero inside the transition curve. At finite $\lam$, as pointed out in~\cite{Faulkner:2008qk}, for any $\mu_q \neq 0$ at any finite temperature, the baryon
charge density is in fact nonzero, given by a dilute Boltzmann gas of quarks.
This immediately raises the question whether the transition along~\eqref{cur} is an
artifact of the infinite $\lam$ limit and will be smoothed into a crossover
at any finite $\lam$. To settle this question requires summing over the worldsheet instantons found in~\cite{Faulkner:2008qk} and will be pursued
in a separate publication.

The plan of the paper is as follows. In section~\ref{sec:2} we discuss
the gravity description of the gauge theory system. In sec~\ref{sec:gen} we review the general gravity setup for describing $\NN=4$ SYM theory coupled to $N_f$ fundamental hypermultiplets at finite temperature.
In sec.~\ref{sec:2b} we review how to introduce a finite baryon chemical potential.
In sec.~\ref{sec:ins} we argue that the phase transition is driven by string worldsheet
instantons.
In sec~\ref{sec:sma} we give a detailed analytically derivation of equation~\eqref{criB} and verify it numerically. In sec.~\ref{sec:ther}
we discuss the thermodynamics of the system which can
be gathered from equation~\eqref{criB}.
In the conclusion sec.~\ref{sec:con} we discuss the connection of
the critical point to
 the first order dissociation transition at $\mu_q=0$.
We also discuss what happens to the phase diagram at finite $\lam$.

\section{Gravity description of the gauge theory} \label{sec:2}

\subsection{General set-up} \label{sec:gen}

At finite temperature, $\NN=4$ SYM theory with a gauge group $SU(N_c)$ can be described by a string theory in
the spacetime of a black hole in AdS$_5 \times S_5$, whose metric can be written as
 \be \label{adsM}
 {ds^2} =  {r^2 \ov R^2} \le(- fdt^2 + d \vec x^2  \ri) + {R^2 \ov r^2} {dr^2 \ov f}  + R^2
 d \Om_5^2
 \ee
where $\vec x = (x_1, x_2, x_3)$ and
 \be
 f = 1-{r_0^4 \ov r^4} \ .
  \ee
$d \Om_5^2$ is the metric on a unit five-sphere $S_5$.
The string coupling $g_s$ and the curvature radius $R$ (in units of $\apr$) are
related to the Yang-Mills coupling $g_{YM}$ and $N_c$ by
 \be
 4 \pi g_{s} = g_{YM}^2, \qquad {R^2 \ov \apr} = \sqrt{\lam}, \qquad \lam = g_{YM}^2 N_c \ .
 \ee
The temperature $T$ of the YM theory
is given by the Hawking temperature of the black hole,
 \be
T = \frac{r_0}{ \pi R^2} \ .
 \ee

One can introduce ``quarks''  to the system by adding to $\NN=4$ SYM theory $N_f$ $\NN=2$ hypermultiplets in the fundamental representation of the gauge group. In the limit of large $N_c$ with $N_f$ finite this can be described in the dual string theory side by adding
$N_f$ D7-branes in the black hole geometry (\ref{adsM}) and to leading order in $N_f/N_c$, the backreaction of the D7-branes on the background geometry can be neglected.
A fundamental ``quark'' in the YM theory can be described by an open string with one end on the D7-branes and the other end on the black hole. Strings corresponding to quarks and anti-quarks have opposite orientations. Open strings with both ends on the D7-branes can be considered as ``bound states'' of a quark and antiquark, thus describing meson-type excitations in the YM theory.

In the limit
 \be
 N_c \to \infty, \qquad \lam \to \infty , \qquad N_f = {\rm finite}
 \ee
which corresponds to the limit $g_s \to 0, {\apr \ov R^2} \to 0$ in the string theory side, the geometric embedding and the dynamics of the D7-branes can be described using the Dirac-Born-Infeld (DBI) action
 \be \label{BI}
 S = -N_f T_7 \int d^{8} \xi \, \sqrt{-\det \le(g_{mn} + 2 \pi \apr F_{mn} \ri)} \ .
 \ee
where $g_{mn}$ denotes the induced metric on the D7-brane and  $T_7 = {1 \ov (2 \pi)^7 g_s \apr^4}$ is the tension of a D7-brane. This will be the limit we work with for most of the paper. We will also consider in the discussion section what happens when one relaxes the $\lam = \infty$ limit, where new interesting physics could emerge.

\begin{figure}[h!]
\centerline{\hbox{\psfig{figure=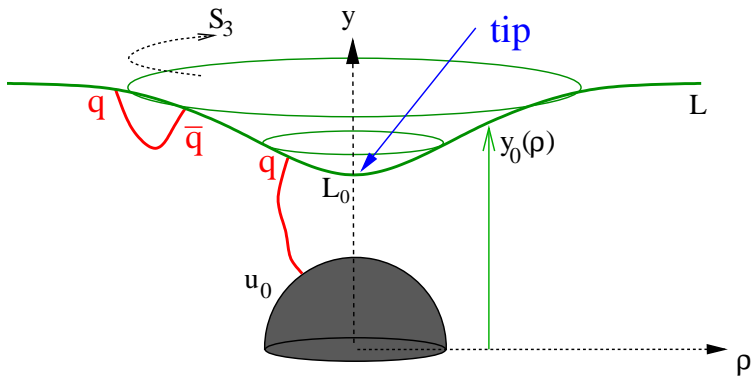,width=10cm
}}}
\vspace{-1cm}
\caption{An embedding of the D7 brane (green) in the $AdS_5 \times S_5$ black hole
geometry which lies entirely outside the black hole. The exact form
of the embedding has been exaggerated to emphasize certain features.
\label{fig:1} }
\end{figure}

To describe the embedding of the D7-branes in~(\ref{adsM}), it is more convenient to introduce a new radial coordinate $u$
defined by
 \be
 {dr^2 \ov  r^2 f(r)} = {du^2 \ov u^2}, \quad  \Longrightarrow \quad   r^2 (u) = {u^4 + u_0^4
 \ov  u^2} \;\; {\rm with} \;\; u_0 \equiv {r_0 \ov \sqrt{2}}
 \ee
in terms of which~(\ref{adsM}) can be written as
  \bea
  \label{eq:met}
  ds^2 & = &   - {r^2 (u) \ov R^2} f (u) dt^2 + {r^2 (u) \ov R^2}d \vec x^2 + {R^2 \ov u^2} ( du^2 + u^2 d \Omega_5^2) \cr
 & = & {u^2 \ov R^2} q(u) \le(-f  dt^2 +  d \vec x^2 \ri) + {R^2 \ov
u^2} \le(d \rho^2 + \rho^2 d \Om_3^2 + dy^2 + y^2 d \phi^2 \ri)  \
 \label{eyr}
 \eea
where
 \be \label{infm}
 u^2 = y^2 + \rho^2, \qquad  f(u)
=\frac{(u^4-u_0^4)^2}{(u^4+ u_0^4)^2}  , \qquad  q(u) \equiv {r^2 (u) \ov u^2} = 1 + {u_0^4 \ov u^4} \ .
\ee
In (\ref{eyr}), we have split the last term of the first line in
terms of  polar coordinates on $\RR^4 \times \RR^2$ with $d \Om_3^2$ denoting the metric on a unit three-sphere. The D7-branes can be chosen to lie along the directions $\xi^\al = (t, \vec x , \Om_3, \rho)$ and using the symmetries of the problem the embedding in the two remaining transverse
directions can be taken as $\phi(\xi^\al) = 0$ and $y(\xi^\al) = y(\rho)$. $y(\rho)$ can be found by solving the equation of motion obtained from the DBI action~\eqref{BI} with the boundary condition
 \be \label{ybd}
 y (\infty) = L = (2 \pi \apr) m_q
 \ee
where $m_q$ can be interpreted as the (bare) mass of the ``quarks''. We denote the resulting embedding function as $y_0 (\rho)$,  which were first obtained numerically in~\cite{Babington:2003vm}.  At a small temperature, the brane lies entirely outside the black hole, as indicated schematically in~Fig.~\ref{fig:1}.
The brane is closest to the black hole at $\rho=0$, where the three-sphere in last term of  (\ref{eyr}) shrinks to a point. Denoting
 \be \label{tip}
 y_0 (\rho=0) =L_0
  \ee
 then the shortest open string connecting the D7-brane to the horizon has a mass in the YM theory
 \be \label{quM1}
 m_q^{(T)}  = {1 \ov 2 \pi \apr} \int_{u_0}^{L_0} dy \, \sqrt{f q}\, \biggr|_{\rho =0}
 \ . 
 \ee
$m_q^{(T)}$ can be interpreted as the effective mass of the ``quarks'' at temperature $T$.
Note that $m_q^{(T)}$ decreases monotonically with $T$, since as we increase the temperature, the black hole becomes bigger and gravity attracts the brane more to the black hole. There exists a temperature $T_d$ 
($=2.166 m_q/\sqrt{\lambda}$), after which the branes fall into the black hole (often called black hole embedding) through a first order phase transition~\cite{Mateos:2006nu,Mateos:2007vn}.
We will be interested in the temperature range smaller or of order $T_d$ in which regime we always have~($\beta = {1 \ov T}$)
 \be \label{qmas}
 \beta m_q^{(T)} \sim O(\sqrt{\lam} ) \ .
 \ee

\subsection{Finite baryon chemical potential} \label{sec:2b}

The system possesses a $U(1)_B$ global baryon symmetry under which ``quarks'' transform nontrivially. The associated conserved current $J^\mu_B$ is dual to the $U(1)$ gauge field $A_\mu$ on the D7-branes. Turning on a chemical
potential $\mu_q$ for the charge density in the boundary theory then corresponds to imposing the following boundary condition\footnote{Also, if the D7-branes fall into the
horizon there will also be a boundary condition that $A_0 = 0$ at the horizon.}
for $A_\mu$
 \be \label{abd}
 A_0 (\rho = \infty) = \mu_q \ .
 \ee
The charged density $n_q$ can be calculated from the bulk theory by
 \be \label{den}
 n_q = \Pi^0 (\rho = \infty)
 \ee
 where $\Pi^0$ is the canonical momentum conjugate to $A_0$ (in terms of $\rho$-slicing) evaluated at the classical solution which satisfies the boundary conditions~\eqref{ybd} and~\eqref{abd}. Again by symmetry we will take $A_0$ to depend only on $\rho$ and the other components of the gauge field will be set to zero.

The DBI action~\eqref{BI} can be written explicitly in terms of $y(\rho), A_0 (\rho)$ as
 \be \label{lad}
 S = - N_f T_7 \int d^8 \xi \, \rho^3 q^{3/2} \left[
fq \left(1+ (y')^2\right) - a'^2 \right]^\ha,
 \ee
 where prime denotes derivative with respect to~$\rho$, and we have introduced
 \be
 a (\rho) = 2 \pi \apr A_0  \ .
 \ee
We will denote the integrand of~\eqref{lad} as $\sL$. Since~\eqref{lad} does not depend on  $a$ (only $a'$)  we have a conserved quantity $\ep$:
\begin{eqnarray}
\label{ee1}
\ep &=& \frac{\partial \sL}{\partial a'}
=  \frac{a' \rho^3 q^{3/2}}{ \left( fq (1+(y')^2)
- a'^2 \right)^\ha } \\
\label{ee2}
a'^2 &=& f q \left(1 + y'^2 \right) \frac{ \ep^2}{
\rho^6 q^3 + \ep^2} \ .
\end{eqnarray}
From~\eqref{den}, we thus find the charge density in the boundary gauge theory is given by
 \be \label{dere}
 n_q = N_f T_7 (2 \pi \apr) (2 \pi^2) \ep
 \ee
 where the factor $2 \pi^2$ comes from the volume of the three sphere.

Since we will be expanding in terms of small density later, it will be convenient to scale coordinates and $\ep$ so that they are dimensionless, i.e.
  \be \label{scel}
 u, y, \rho \to L_0 (u, y, \rho), \quad \ep \to L_0^3 \ep, \quad a(\rho) \to L_0 a (\rho)
 \ee
where $L_0$ is the location of the tip of the brane~\eqref{tip} {\it before}
turning on a chemical potential\footnote{One can also choose to normalize them using $L$~\eqref{ybd} which is more directly related to the field theory mass. But in our calculation below using $L_0$ is slightly more convenient. $L_0$ is fixed once the ratio $m_q/T$ is given. Another alternative is to use the temperature but that will make the zero temperature limit more subtle.}.
From now on, $u,y,\rho, \ep, a(\rho)$ are all dimensionless. After the scaling~\eqref{infm} become
 \be
 \label{rfdr}
 u^2 = y^2 + \rho^2, \qquad q(u) \equiv 1 + {\eta^4 \ov u^4} , \quad f \equiv \le({u^4 -\eta^4 \ov u^4 +\eta^4} \ri)^2, \qquad
  \eta \equiv  {u_0 \ov L_0} < 1  \ .
 \ee
The boundary conditions for $y(\rho)$ and $a (\rho)$ now are
   \be \label{bds}
 y (\infty) = {L \ov L_0}, \qquad a(\infty) =  {2 \pi \apr \ov L_0} {\mu_q } \
 \ee
and equation~\eqref{quM1} becomes
 \be \label{quM}
 m_q^{(T)}  = {L_0 \ov 2 \pi \apr} \int_{\eta}^{1} dy \, \sqrt{f q}\, \biggr|_{\rho =0}
 \ . 
 T\ee

\begin{figure}[h!]
\centerline{\hbox{\psfig{figure=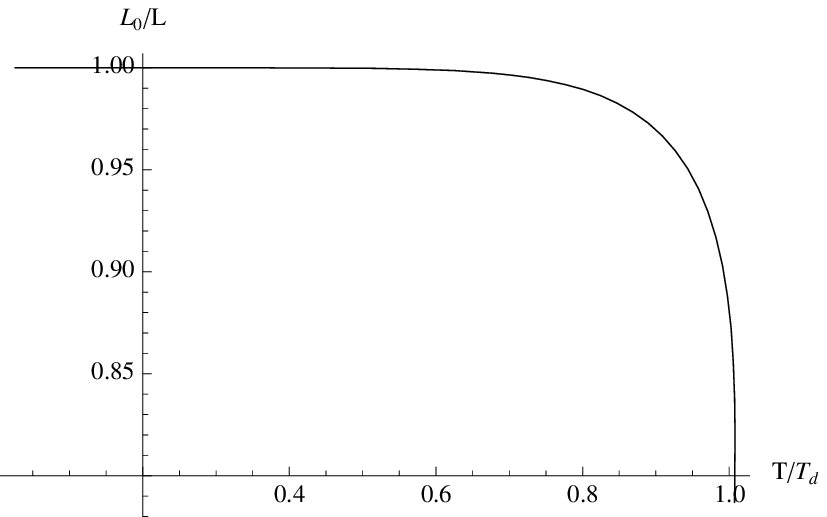,width=7.5cm
}} \hbox{\psfig{figure=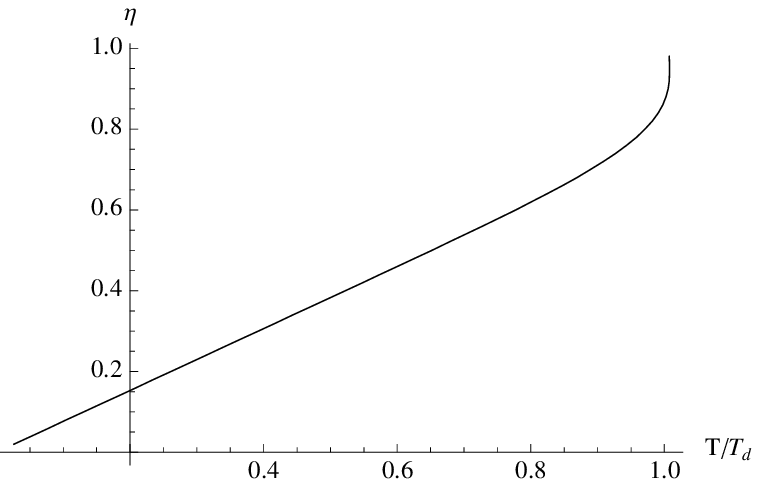,width=7.5cm
}} } 
\caption{$L_0/L$ and $\eta$ for the Minkowski solution ($\ep = 0$) 
as a function of $T/T_d$. 
\label{fig:Leta}}
\end{figure}

To obtain the equation of motion for $y$ it is convenient to perform a Legendre
transformation on~\eqref{lad} to express it in terms of $\ep$. The transformed action is
\be
\label{eq:tosolve}
\mathcal{H} \equiv \ep a' +\sL = \sqrt{\ep^2 + \rho^6 q^3} \sqrt{f q (1+y'^2)}
\ee
which  leads to equation of motion
 \be
  {y'' \ov 1+y'^2} + {3 y' \ov \rho} + {8 \ov u^2} {\rho y' - y \ov u^8 \eta^{-8} -1}- {\ep^2 \ov \ep^2 + \rho^6 q^3} \le[{3 y' \ov \rho}
 - {6 \ov u^2} {\rho y' - y \ov 1 + u^4 \eta^{-4} }  \ri] =0 \ .
 \label{eom}
 \ee
When $\ep =0$ (i.e. zero density), equation (\ref{eom}) becomes
 \be \label{zero}
  {y'' \ov 1+y'^2} + {3 y' \ov \rho} + {8 \ov u^2} {\rho y' - y \ov u^8 \eta^{-8} -1} =0
  \ee
whose solution is given by $y_0 (\rho)$ described earlier around equation~\eqref{tip}.\footnote{As stated earlier we will only consider $T < T_d$ for which there is only one Minkowski embedding solution.} In Fig.~\ref{fig:Leta} 
we plot how $L_0/L$ and $\eta$ change with temperature for an embedding governed by $y_0 (\rho)$.

\subsection{Phase transition driven by string worldsheet instantons} \label{sec:ins}

For {\it any} value of $\mu_q$ (and $T<T_d$), one can always have the following solution to equations of motion~\eqref{ee2} and~\eqref{eom} with
 \be \label{triv}
 A_0 = \mu_q = {\rm const} , \qquad \ep =0, \qquad y (\rho) = y_0 (\rho) \
 \ee
for which~\eqref{eom} reduces to~\eqref{zero}. This solution is somewhat peculiar, since it implies that the quark charge density $n_q$ is zero even at a finite temperature and finite chemical potential $\mu_q$. This appears to contradict with field theory expectation that the density for quark and anti-quark should be given (at low density) by the Boltzmann distribution $n_\pm \sim e^{-\beta (m_q^{(T)} \pm  \mu_q)}$ which gives a nonzero net charge density $n_q = n_+ - n_- \neq 0$ for a nonzero $\mu_q$. There is in fact  no contradiction, since~\eqref{triv} is the result in the $\lam = \infty$ limit (supergravity limit) and in this limit due to~\eqref{qmas},
 \be
n_\pm \sim n_q \sim e^{-\sqrt{\lam}}, \quad {\rm for} \quad
\mu_q < m_q^{(T)} \ .
\ee
The charge density is thus exponentially small in the large $\lam$ limit and not visible to any order in the $1/\sqrt{\lam}$ (or $\apr$) expansion.

\begin{figure}[h!]
\centerline{\hbox{\psfig{figure=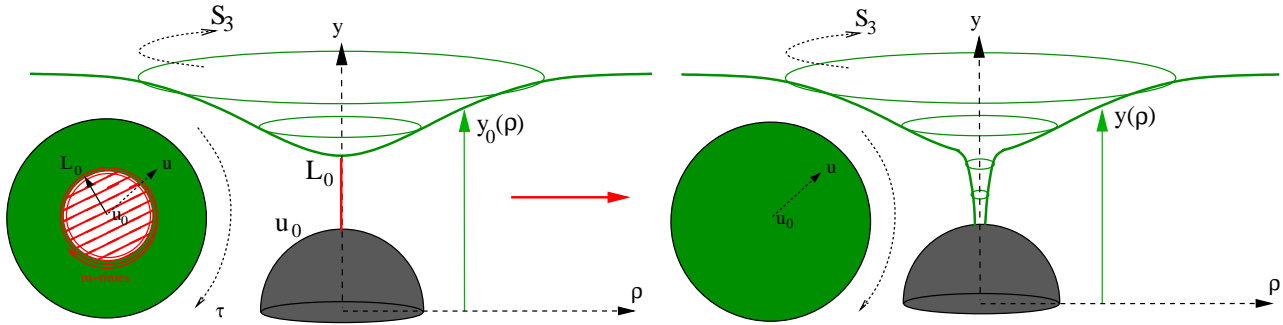,width=17cm
}}}
\vspace{-1cm}
\caption{Left plot: For Minkowski type embedding~\eqref{triv}, there exist
worldsheet instantons which correspond to semi-classical strings stretching
between the tip of the D7-branes to the black hole horizon and winding around the
Euclidean time direction represented as the angle on the
inset disk. The radial direction on the disk is the same
as the radial direction in the $y-\rho$ plane.
Right plot: for $\mu_q > m_q^{(T)}$
instantons dominate over the DBI contributions and will condense to form a genuine
neck between the branes and the horizon, i.e. one should have a black hole type embedding.
\label{fig:2} }
\end{figure}

In~\cite{Faulkner:2008qk}, it was found for the embedding~\eqref{triv} there are non-perturbative open string worldsheet instanton corrections to the DBI action~\eqref{BI} which accounts for the exponentially small quark density. More explicitly, the instantons
are given by open strings stretching between the
tip of the D7-branes and the black hole horizon and winding around the Euclidean
time direction, as indicated in the left plot of Fig.~\ref{fig:2}. They are classified a winding number $n$. From the spacetime point of view, these instantons generate tiny virtual necks which connect the tip of the branes to the black hole horizon. The total Euclidean action for the D7-branes, which gives the thermodynamic potential of the boundary gauge theory in the grand canonical ensemble, can be written as
 \be \label{inst}
 S = S_{DBI} + \sum_{n = \pm 1, \pm 2 , \cdots} D_n \exp \le(-|n| \beta (m_q^{(T)} - \sgn[n] \mu_q)\ri) + \cdots
 \ee
where $n$ sums over the instanton contributions, $D_n$ arises from the worldsheet determinant for each instanton\footnote{It also includes possible integrations over instanton moduli space.}, and $\cdots$ denotes
other perturbative $1/\sqrt{\lam}$ (or $\apr$) corrections. These string worldsheet instantons have a simple interpretation in the boundary gauge theory as representing thermal medium quarks. In particular, the $n = \pm 1$ terms in~\eqref{inst} are precisely what one expects of a dilute Boltzmann gas of quarks and anti-quarks.\footnote{Higher $n$ contributions should encode corrections due to Bose-Einstein or Fermi-Dirac statistics and
other corrections due to interactions.}

Due to~\eqref{qmas}, when $\mu_q < m_q^{(T)}$ the instanton contributions in~\eqref{inst} are exponentially small in $\sqrt{\lam}$ compared with the DBI action $S_{DBI}$. But for
 \be \label{rab}
 \mu_q \geq m_q^{(T)}
 \ee
 the instanton sum will be exponentially large compared with the DBI action and the solution~\eqref{triv} can no longer be trusted. In particular, since the instanton contributions are dominating, we expect them to ``condense'' and create a genuine neck between the brane and the black hole. This implies in the range~\eqref{rab} a new solution to the DBI action with the branes going into the black hole should emerge, as indicated
 in the right plot of Fig.~\ref{fig:2}. Thus in the infinite $\lam$ limit we have a phase transition at $\mu_q = m_q^{(T)}$ where the Minkowski-type embedding~\eqref{triv} goes over to a black-hole-type embedding. The transition also has a simple interpretation
from the gauge theory point of view; when $\mu_q$ satisfies~\eqref{rab}, the quarks have negative free energies and will thus condense and generate a finite charge density.

The possibility for such a phase transition has been studied before in~\cite{Mateos:2007vc,Ghoroku:2007re,Nakamura:2007nx}, 
where a single\footnote{at a temperature not too high} black hole embedding solution was found for $\mu_q > m_q^{(T)}$ and it was concluded 
in \cite{Mateos:2007vc} that the transition was {\it first order}. One reason for the conclusion was that it appeared that the Minkowski embedding solution~\eqref{triv} was still valid for $\mu_q > m_q^{(T)}$ and thus the transition appeared to be a change of dominance between different solutions. As we discussed above, the Minkowski embedding solution should be {\it replaced} by a black hole embedding solution
in the parameter region~\eqref{rab}. Thus if there is only a single black hole embedding solution for $\mu_q > m_q^{(T)}$ the transition is most likely to be continuous.
Indeed we will show in section~\ref{sec:ther}, the transition is {\it third order}
for temperature not too high, but then becomes first order through a {\it tricritical point}.\footnote{Beyond the tricritical point it becomes possible to have multiple
black hole embedding solutions for a given $\mu_q$. See section~\ref{sec:ther}.}

\subsection{Black hole embedding at finite density}

As discussed in last subsection, for $\mu_q \geq m_q^{(T)}$ we expect a black hole type embedding for D7-branes in the gravity side and a nonzero quark charge density on the gauge theory side. That the quark density for a black hole type embedding should be zero can be seen as follows. For a black hole type embedding, one should impose an additional boundary condition $A_0 =0$ for the gauge field at the horizon in order to ensure the regularity of the solution. From~\eqref{abd} this implies that $A_0$ should evolve nontrivially from the boundary to the horizon, i.e. there is a nontrivial electric field in the radial direction on the branes, which  from~\eqref{den} in turn implies $n_q  \neq 0$ (and thus $\ep \neq 0$).
Thus we should now consider equation~\eqref{eom} with $\ep \neq 0$. Conversely it has also been argued previously in~\cite{Kobayashi:2006sb} that for any $\ep \neq 0$, only black hole type of embeddings are allowed since for a Minkowski type embedding it is not possible to have a non-singular distribution of a finite charge density on the branes\footnote{See also~\cite{Karch:2007br} for a discussion at zero temperature}.

Let us now consider the boundary conditions for $y (\rho)$ at the horizon
for a black hole type embedding. We will denote the point that the brane enters the horizon $\rho_c$ and $y_c \equiv y (\rho_c) $.
Note that
\be \label{hbd1}
y_c^2 + \rho_c^2 = u_0^2
\ee
and in order for the the third term in~\eqref{eom} to
be nonsingular at the horizon, we will also need that
 \be \label{hbd2}
 y' (\rho_c) = {y_c \ov \rho_c}
 \ee
i.e. the brane should be perpendicular to the horizon. The precise of value of $\rho_c$ is determined by the boundary condition~\eqref{bds}.

For regularity we will also require the gauge potential vanish at the horizon\footnote{Otherwise the one-form $A_0 dt$ will be singular at the horizon since the norm for ${\p \ov \p t}$ vanishes there.}, i.e.~$a (\rho_c) =0$. From equation~\eqref{ee2}, we can express the chemical potential in
  terms of density $\ep$ as
   \be \label{mu}
\mu_q = {L_0 \ov 2 \pi \apr} \ep\int_{\rho_c}^\infty d \rho \, {\sqrt{fq (1+y'^2)} \ov \sqrt{\ep^2 + \rho^6 q^3}} \ .
 \ee
The main technical task of the paper is to determine the behavior of $\mu_q$ in the limit of small~$\ep$, which will be carried out in next section.  Once the expansion of $\mu_q$ in
terms of $\ep$ is known, we will be able to determine the order of the phase transition
and other thermodynamic properties of the system. This will be carried out in section~\ref{sec:ther}.

\section{Small density expansion of the chemical potential} \label{sec:sma}

In this section we study the behavior of~\eqref{mu} in the small $\ep$ limit.  Expanding~\eqref{mu} in small $\ep$ is somewhat
complicated since the solution $y (\rho)$ depends nontrivially on $\ep$. The expansion of $y (\rho)$ in terms of small $\ep$ is also involved; one cannot treat the $\ep$-dependent term in (\ref{eom}) as a small perturbation of~\eqref{zero} uniformly for all values of $\rho$ since the term becomes of $O(1)$ for sufficiently small $\rho$ (when $\rho^6 q^3 \sim O(\ep^2)$).

The main result of the section is equation~\eqref{muF}.  Readers who are not interested in the detailed derivation can go directly to~\eqref{muF} and subsequent discussions.

\subsection{Expansion of the solution}

Since the perturbation in $\ep$ is not uniform in $\rho$ one needs to divide the $\rho$ axis into different regions, treating the perturbations in each region separately, and matching them together at the end. For our purpose, it turns out enough to split the $\rho$-axis into two regions:
\begin{itemize}
\item \textbf{Inner} $\rho = \eps^{\ha} \sig$ for $\sig_c < \sig <\Lambda$
\item \textbf{Outer} $\rho_\Lambda <\rho<\infty$
\end{itemize}
where
\be
\rho_c = \ep^\ha \sig_c, \qquad \rho_\Lambda =  \ep^\ha \Lam  \ .
\ee
The reason for the choice of scaling in the inner region can be seen by letting $\rho = \ep^\al \sig$ in~\eqref{eom} and one finds that for $\ep = \ha$ a nontrivial scaling limit exists, which results differential equation in (\ref{e4}) below. Also see Appendix B for
a discussion of the $T=0$ solution where this scaling is evident.

We will consider the limit
 \be
 \ep \to 0, \qquad \sig = {\rm finite} , \qquad \Lam \to \infty, \qquad \rho_\Lam \to 0 \ .
 \ee
We expand the solution in the inner and outer regions as\footnote{While it is not a priori obvious the expansion parameter should be $\eps$ and not some other power, the expansions below do yield consistent results.}
 \bea \label{in}
\yI(\sig) & = & Y_0 (\sig) + \eps Y_1 (\sig) + \eps^2 Y_2 (\sig) + \cdots \\
\label{out}
\yO(\rho) & =& y_0 (\rho) + \eps y_1 (\rho) + \eps^2 y_2 (\rho) + \cdots \ .
\eea
Plugging~\eqref{in} and~\eqref{out} respectively into~\eqref{eom} and expanding the
equation in $\ep$, we obtain a series of differential equations for various functions in~\eqref{in} and~\eqref{out}.
These equations are rather complicated and not exactly solvable. We will work out the behavior of $Y_0 (\sig), Y_1 (\sig)$ in the large $\sig$ limit
and $y_0 (\rho), y_1 (\rho) $ in the small $\rho$ limit and match them in the overlapping region. Fortunately it turns out this is enough
to find the leading expansion of $\mu_q$ in the small $\ep$ limit analytically.

Let us first examine the outer region. $y_0$ is simply the solution to~\eqref{zero} which describes the embedding at zero density. It satisfies the boundary condition~\eqref{bds} at the boundary and near $\rho =0$ has the following expansion
 \be
 \label{outer}
    y_0 (\rho) = 1 + {\rho^2 \ov \eta^{-8} -1} - {\eta^8 (5+5\eta^8 - 3 \eta^{16}) \ov 3 (1-\eta^8)^3} \rho^4 + O(\rho^6) \ .
    \ee
$y_1$ satisfies a homogeneous
linear equation obtained by linearizing (\ref{zero}). For small $\rho$ it has an expansion,
\be \label{ex1}
y_1 =  b_{-1}  \rho^{-2} 
+  b_0 \log \rho + b_1  
+ O(\rho)
 \ee
where $ b_{-1}$ and $b_1$ are integration constants and
 \be \label{re1}
 b_0 =  - \frac{  28 \eta^8 }{(1 - \eta^4)^2
(1 + \eta^4)^2 } b_{-1} \ .
 \ee
Note that the first two terms diverge in the $\rho \to 0$ limit. The coefficient $b_{-1}$ of the quadratically divergent term will be fixed
later by matching with the inner solution. $b_1$ should be determined by requiring
$y_1 (\rho \to \infty) \to 0$ since $y_0$ already satisfies the required boundary condition~\eqref{bds} there.

Now we look at the inner region. $Y_0 (\sig)$ satisfies a second order \emph{non-linear}
differential equation (primes below denotes derivative w.r.t. $\sig$)
 \be \label{e4}
{Y_0'' \over Y_0'^2} +  {3 \left(1+ \frac{\eta^4}{Y_0^4} \right)^3 \sig^5 Y_0'}
  + {8 \over Y_0^2} {\sig Y_0' - Y_0 \over Y_0^8 \eta^{-8} -1}
  + {6 \over Y_0^2} {\sig Y_0' - Y_0 \over 1 + Y_0^4 \eta^{-4} } =0
 \ee
From~\eqref{hbd1} and~\eqref{hbd2} the boundary condition at the horizon is
\begin{equation} \label{bd1}
Y_0 (\sig_0) = \eta, \qquad Y'_0 (\sig_0) = {\eta \ov \sig_0}
\end{equation}
where $\sig_0$ is an integration constant. $\sig_0$ is the zero-th order term in the expansion in $\ep$ of $\sig_c = \rho_c/\ep^{1/2}$ (where
again it is a non trivial fact that this expansion begins at
$\mathcal{O}(1)$ in $\ep$)
\be \label{hex}
\sig_c = \sig_0 + \ep \, \sig_1 + \cdots
\ee
and can be determined
by the boundary condition at infinity,
\begin{equation}
Y_0 (\sig) \rightarrow 1 \qquad {\rm as}\qquad
\sig \rightarrow \infty \ .
\end{equation}
The above condition is fixed by comparing with the leading term in~\eqref{outer}.
At large $\sig$, $Y_0$ has the expansion
 \bea
 Y_0 (\sig) && = 1 - \frac{1}{2 (1+\eta^4)^{3/2}} \sig^{-2} +
 \frac{\eta^4( - 3 + 5 \eta^4)}{4 ( -1 + \eta^4)(1 + \eta^4)^4} \sig^{-4} \cr
 && \qquad + \quad \frac{\eta^4 ( 27 - 131 \eta^4 + 153 \eta^8 - 105 \eta^{12})}
{24 (-1 + \eta^4)^2 ( 1 + \eta^4)^{13/2} } \sig^{-6} + O (\sig^{-8}) + O(\sig^{-8} \log \sig)  \label{y0E}
\eea
Note that the coefficient before $\sig^{-8}$ is an integration constant which can be fixed by~\eqref{bd1}. We will not need its value below.

 $Y_1 (\sig)$ satisfies a second order \emph{linear} differential
equation with coefficients that depend on $Y_0 (\sig)$.
It is not homogeneous and rather complicated. We will not write it here.
For large $\sig$ we find the expansion
\begin{equation}
\label{eq:inassym1}
Y_1 (\sig) =  a_{-1}
\sig^2 + a_0 \log \sig + a_1 + O (\sig^{-1})  
\end{equation}
where $a_1$ is an integration constant (the other integration
constant presumably appears at higher order in the expansion) and
 \be \label{re2}
 a_{-1} = \frac{\eta^8}{1 - \eta^8}, \qquad
 a_0 =   \frac{  14 \eta^8  }{(1 - \eta^4)^2
(1 + \eta^4)^{7 \ov 2} } \ .
 \ee
Note the first two terms of~\eqref{eq:inassym1} are divergent as $\sig \to \infty$.
{The  constant $a_1$ can be determined by
matching with the solution of the outer region.
The \emph{other} boundary condition is
determined by
the regularity condition~(\ref{hbd1}-\ref{hbd2})
(expanded to this order) at the horizon,
this gives only one boundary condition on $Y_1$ and its
derivative because  the first order correction $\sig_1$
in~\eqref{hex} is a free parameter.
This parameter $\sigma_1$ would then be determined
by the resulting solution.

We will now determine various integration constants above by matching the solutions
in the inner and outer regions around $\rho_\Lam$. Note given the relation $\rho = \ep^\ha \sig$, the $\ep$ expansion in the outer region is reshuffled compared to that in the inner region. Comparing the first term in~\eqref{ex1} to the second term of~\eqref{y0E} we find that
 \be \label{par1}
 b_{-1}= -\frac{1}{2 (1+\eta^4)^{3/2}}, \qquad 
 b_0 = \frac{14 \eta^8 }{(1 - \eta^4)^2
(1 + \eta^4)^{7 \ov 2} }
 \ee
where we have used~\eqref{re1} in obtaining the second equation.
From~\eqref{par1} and~\eqref{re2} we see that the logarithmic term in~\eqref{ex1} precisely match with that in~\eqref{eq:inassym1}; this is a nontrivial self-consistency
 check of our expansion.  It now remains to match the constant terms in~\eqref{ex1} and~\eqref{eq:inassym1}. Given the difference in the argument of logarithmic term in~\eqref{ex1} and~\eqref{eq:inassym1}, we thus find that
  \be \label{inC1}
  a_1 = b_1 + \ha b_0 \log \ep = b_1 +  \frac{7 \eta^8 }{(1 - \eta^4)^2
(1 + \eta^4)^{7 \ov 2} } \log \ep  \ .
  \ee
 As remarked below~\eqref{re1}, $b_1$ is determined by the boundary condition $y_1 (\infty) =0$ and since the equation for $y_1$ is independent of $\ep$, so is $b_1$.\footnote{Determining $b_1$ requires solving the  two-point boundary value problem for $y_1$. 
 }
 Thus we see from~\eqref{inC1} that the integration constant $a_1$ for inner
 solution $Y_1$ now contains a $\log \ep$ piece!\footnote{This in turn implies that $\sig_1$ in~\eqref{hex} also contains $\log \ep$.} This will be important in our determination of the leading order behavior for $\mu_q$ below.

As a check on our above results we can make sure that
the exact (numerical) solution at small densities matches
well onto our expansions in the two different regions.
This is demonstrated in Fig.~(\ref{fig:3}). The
agreement is very precise suggesting the two regions
we have used are sufficient. Higher order
corrections in the two regions would appear
as divergences toward the exact solution.

\begin{figure}[h!]
\centerline{\hbox{\psfig{figure=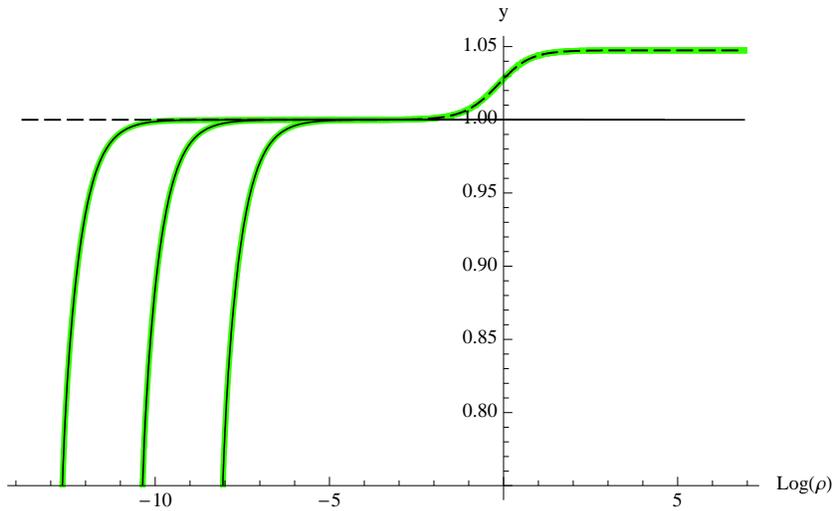,width=11cm
}}}
\caption{ The exact embeddings for $\epsilon = 10^{-11}, 10^{-9}, 10^{-7}$
shown in green. To clearly show the two regions
we have plotted $y(\rho)$ as a function of $\log(\rho)$.
These are compared to the zeroth order inner (\emph{solid})
and outer (\emph{dashed}) solutions.
The inner solution $Y_0(\sigma)$ is a function
of $\sigma = \rho \epsilon^{-1/2}$
so for the various fixed densities the curve
is simply shifted along the $\log(\rho)$ axis. 
The agreement between the three curves
on the overlapping regions is very precise.
\label{fig:3} }
\end{figure}

\subsection{Expansion of the chemical potential} \label{sec:epr}

We now look at the small $\ep$ expansion of~\eqref{mu}. We will give the main steps and
final results, leaving further details to Appendix~\ref{app:C}.

The $\mu_q$ integral~\eqref{mu} can be split into integrals over the inner and outer regions
 \be
 \mu_q = \int_{\rho_c}^{\rho_\Lam} (\cdots) + \int_{\rho_
 \Lam}^{\infty} (\cdots) \equiv \mu_{\sI} + \mu_{\sO} \ .
 \ee
$\mu_{\sI}$ and $\mu_{\sO}$ can now be expanded in terms of $\ep$ using~\eqref{in} and~\eqref{out}. The outer region contribution $\mu_{\sO}$ starts with order $\ep$, i.e.
\be
  \mu_\sO = \ep \, \mu_{\sO}^{(1)} + O(\ep^2)
  \ee
and for $\mu_{\sO}^{(1)}$ only $y_{0}$ is needed. Using the small $\rho$ expansion of $y_{0}$~\eqref{outer} we find that the integral for $\mu_{\sO}^{(1)}$ contains a quadratic and a logarithmic divergent term in the limit $\rho_\Lam \to 0$,
  \be \label{oo1}
 \mu_{\sO}^{(1)} = {L_0 \ov 2 \pi \apr} \le( \frac{ 1- \eta^4}{2 ( 1 + \eta^4)^2 \rho_\Lambda^2 }
- \frac{ 2  \eta^4 (3 - \eta^4 + 3 \eta^8)}{(1 - \eta^4)(1 +\eta^4)^4}
\log(\rho_\Lambda) + K_\sO (\eta, L/L_0) \ri)
 \ee
where $K_\sO$ denotes the part which remains finite in the limit $\rho_\Lam \to 0$
$K_\sO$, whose explicit expression is given in~Appendix~\ref{app:C}, depends on the full solution $y_0 (\rho)$ and can only be evaluated numerically.

The inner region contribution also contains an $O(1)$ piece
 \be
 \mu_{\sI} = \mu_{\sI}^{(0)} + \ep \mu_{\sI}^{(1)} + \cdots
 \ee
with (recall that $\sig = \ep^{-\ha} \rho$)
 \be
 \mu_{\sI}^{(0)} ={L_0 \ov 2 \pi \apr} \int_{\sig_c}^\Lam d \sig \,  Y_0' (\sig) \, {Y_0^4 - \eta^4 \ov Y_0^2 \sqrt{Y_0^4 + \eta^4}} = \int_{\sig_c}^\infty (\cdots) - \int_{\Lam}^\infty (\cdots)
 \ee
where in the second equality we have separated the expression into two pieces by
splitting the integral. Now changing the integration variable of the first integral into $Y_0$ and comparing it with~\eqref{quM}, we find that it is precisely $m_q^{(T)}$. The second term can be evaluated in power series of $1/\Lam$ by using the expression~\eqref{y0E} for $Y_0$ at large $\sig$, leading to
 \be \label{oo2}
 \mu_\sI^{(0)} = m_q^{(T)} - {L_0 \ov 2 \pi \apr} \frac{(1 - \eta^4)}{2 (1 + \eta^4)^2 \Lambda^2} + \mathcal{O}(\Lambda^{-4}) \ .
 \ee
To evaluate $\mu_{\sI}^{(1)}$ one also needs $Y_1$. Using the expansion~\eqref{eq:inassym1} of $Y_0$ and $Y_1$ for large
$\sig$, we find that the integral for $\mu_{\sI}^{(1)}$
 contains a logarithmic divergence (in the $\Lam \to \infty$ limit) from the upper end of the integral and can be written as
 \be \label{oo3}
 \mu_\sI^{(1)} = {L_0 \ov 2 \pi \apr} \le(\frac{2 \eta^4 ( 3 -  \eta^4 + 3 \eta^8)}{(1- \eta^4) (1+ \eta^4)^4}
\log(\Lambda) +  {7 \eta^8 \log \ep \ov (1 - \eta^4)(1+\eta^4)^4} + K_\sI (\eta, L/L_0) \ri) \ .
\ee
Note that in the finite part we have isolated a piece proportional to $\log \ep$ which comes from the integration constant $a_1$ of $Y_1$ through equation~\eqref{inC1} (see Appendix~\ref{app:C} for details). $K_\sI$ is finite and independent of $\ep$. Its explicit expression is given in~\ref{app:C} and can only be evaluated numerically.

Now adding~\eqref{oo1} and~\eqref{oo2},~\eqref{oo3} together we find that
the second term in~\eqref{oo2} precisely cancels the quadratic divergence in~\eqref{oo1} and  the logarithmic divergence in~\eqref{oo3} precisely cancels with that in~\eqref{oo1}, leaving a finite piece proportional to $\ep \log \ep$. We thus find that
\be \label{muF}
 \mu = m_q^{(T)} - B(T) \ep \log \ep +  A(T) \ep + O(\ep^2)
 \ee
where
 \be \label{a1}
 B(T) = B_1 (T) + B_2 (T) = {L_0 \ov 2 \pi \apr} \frac{ \eta^4 \left(
3 - 8 \eta^4 + 3 \eta^8 \right)}{(1 - \eta^4)(1 +\eta^4)^4}, \qquad
 A (T) =  {L_0 \ov 2 \pi \apr} \le(K_\sO + K_\sI \ri) \
 \ee
 and
 \be \label{a2}
 B_1 (T) = {L_0 \ov 2 \pi \apr} \frac{ \eta^4 ( 3 -  \eta^4 + 3 \eta^8)}{(1- \eta^4) (1+ \eta^4)^4}, \qquad B_2 (T) =- {L_0 \ov 2 \pi \apr}  {7 \eta^8 \ov (1 - \eta^4)(1+\eta^4)^4}
 \ee
 Note that in writing~\eqref{a1} and~\eqref{a2} we have emphasized that the coefficient
 of $\ep \log \ep$ receives contributions from two different sources; $B_1$ from cancellation of the logarithmic divergence and $B_2$ from the finite piece in~\eqref{oo3}.
The expressions for $K_\sO$ and $K_\sI$ are given in Appendix~\ref{app:C}, which can only be evaluated numerically.  We also note that in the $\ep \to 0$ limit it is the $\ep \log \ep$ which dominates and whose coefficient we have determined exactly. Note that $L_0$, which gives of the location of the tip of the D7-branes in the absence of a charged density~\eqref{tip}, is also temperature-dependent.

\begin{figure}[h!]
\centerline{\hbox{\psfig{figure=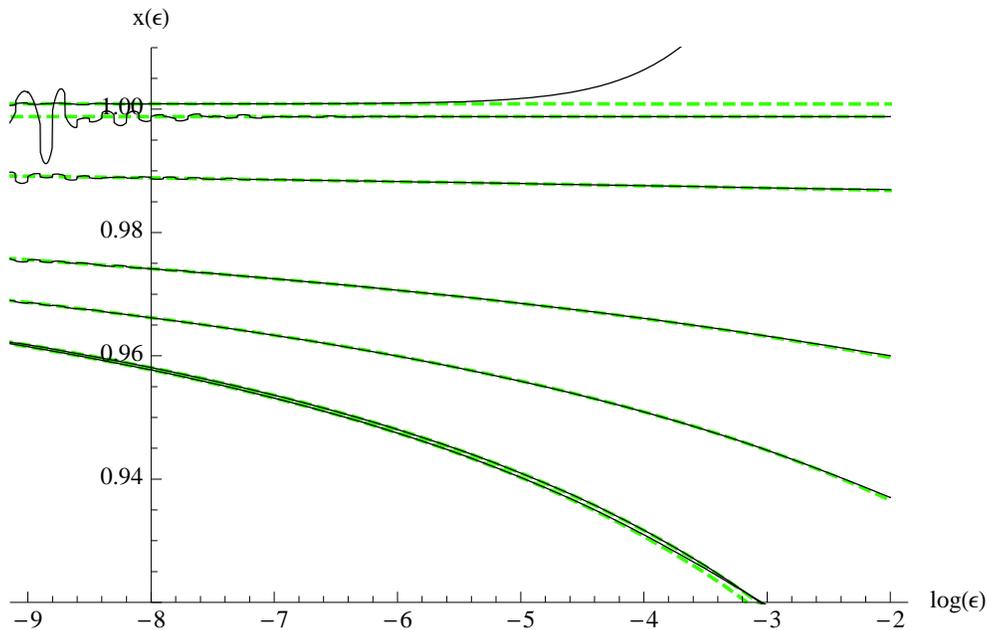,width=13cm
}}}
\caption{Plots of the exponent $x(\ep)$ as defined
in~\eqref{xeq1}, for various values of the temperature.
At zero temperature
the exponent is simply $1$ and all
these curves should eventually limit to $1$
at small enough densities. The slow running
of this exponent is a consequence of the $\log\ep$ behavior
of the chemical potential.
The dashed line is a (1 parameter) fit to the numerical
curve using~\eqref{xeq2}.
\label{fig:4}
}
\end{figure}

Again we can use exact (numerical) solutions
for small densities to check our analysis.
The $\log \ep$ dependence is not easy to see, so we must push our
analysis to very small densities with a wide
range of densities. Then the best way to extract
the behavior is to plot the function $x(\ep)$  defined by,
\bea \label{xeq1}
x(\ep) &=& \frac{ d \log (\mu_q - m_q^{(T)})}{d \log \ep} \\
 & \approx &  \frac{A(T) - B(T) ( \log \ep + 1)}{A(T) - B(T) \log \ep }
 \label{xeq2}
\eea
where for the last equality we have used our derived
small $\ep$ expression
(\ref{muF}).
We can fit this later form to the numerical result and find
the ratio $A(T)/B(T)$, after which we can use the overall
normalization of $\mu_q - m_q^{(T)}$ to fix $A(T)$ and $B(T)$
separately. This agreement in the form of the solution is shown in
Fig.~\ref{fig:4}.

\subsection{Behavior of $B(T)$ and $A(T)$}

\begin{figure}[h!]
\centerline{\hbox{\psfig{figure=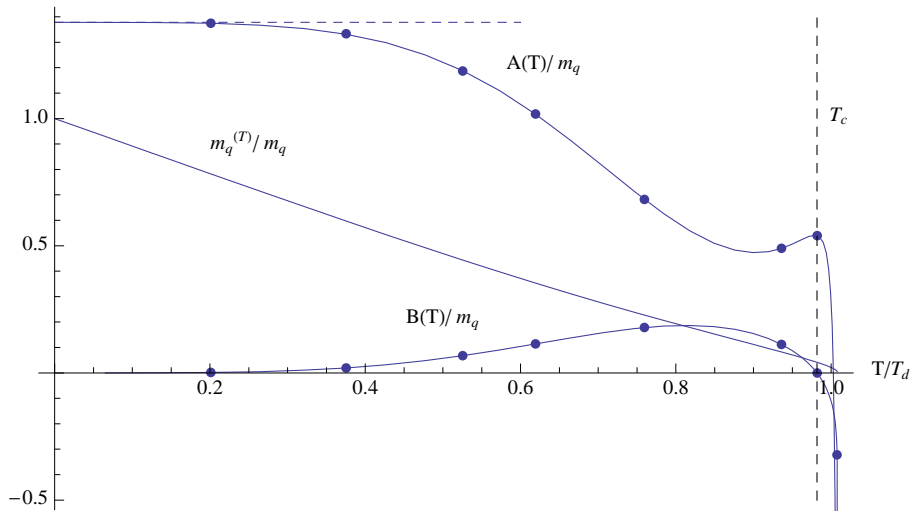,width=12cm
}}}
\caption{The quantities in (\ref{muF}) are plotted
as a function of $T/T_d$. The dots represent the values
of $A(T)$ and $B(T)$ obtained through fitting the
numerical results to the
form of the small density expansion. The actual
curves come from the results of section III (which also
require numerics to calculate $A(T)$.) The consistency
is gratifying.
\label{fig:5}}
\end{figure}

Let us now examine the behavior of $B(T)$ and $A(T)$ as we vary $T$.
In the $T \to 0$ (i.e. $\eta \to 0$) limit, from~\eqref{a1}
 \be
m_q{(T)} = m_q, \qquad A(T) = m_q \frac{\kappa^3}{2} , \qquad  B(T) = 0
 \ee
which reduces to the correct zero temperature behavior (\ref{eq:zt}) in Appendix~\ref{app:zerot}. In the above equation $\kappa$ is a number defined by the integral~$\kappa = \int_0^\infty dx/(x^6 + 1)$. For general temperature $T$,
the behavior of $A(T), B(T)$ and $m_q^{(T)}$ are plotted in
Fig.~\ref{fig:5}.  Note that in obtaining Fig.~\ref{fig:5} we have converted the dependence on $\eta$ into $T$ and also the temperature dependence of $L_0$
using the relations demonstrated in Fig.~\ref{fig:Leta}.

Note that $B(T)$ is positive at small temperature and becomes negative for
\begin{eqnarray} \label{eq:Tc}
\eta > \eta_c, \qquad \eta_c = ({1 \ov 3} (4 - \sqrt{7}) )^{1 \ov 4}  , \quad {\rm i.e.} \quad
T > T_c , \qquad T_c \approx .982 T_d, \qquad B(T_c ) =0
\end{eqnarray}
Also note that $A(T)$ becomes negative at some temperature higher than $T_c$. Furthermore it can be shown that $A(T)$ diverges at a temperature $T_m$ ($T_m 
1.008 T_d$), beyond which Minkowski embedding no longer exists; clearly our analysis does not apply since there is no zero density solution about which to perturb in $\ep$.

\section{Thermodynamics} \label{sec:ther}

From our small $\ep$ analysis we can now extract some important
aspects of the thermodynamics of this gauge theory.
We will mostly work in the grand canonical ensemble, where
we fix $T,\mu_q$ and use the pressure $P(\mu_q,T)$ as the appropriate
thermodynamic potential.

\subsection{A third order phase transition for $0<T<T_c$}

As noted above $B(T)>0$ for $0<T<T_c$. In this case
~\eqref{muF} may be inverted to find
$\ep(\mu_q)$ which is then a single-valued function of $\mu_q$
for $\mu_q > m_q^{(T)}$.
In other words there is only a single black hole
embedding for a given fixed $\mu_q > m_q^{(T)}$.
Thus for $B(T)>0$ we conclude there 
is a continuous phase transition at $\mu_c = m_q^{(T)}$.
For $\mu_q < m_q^{(T)}$ we have a single Minkowski embedding with
$\ep = 0$.

Examining derivatives of the pressure on either side
of $m_q^{(T)}$ one finds for $\mu_q<m_q^{(T)}$,
 \be
  P (T, \mu) = P_0 (T), \qquad n_q = {\p P_0 \ov \p \mu_q} = 0, \qquad
   {\p^n P_0 \ov \p \mu_q^n} = 0, \quad {\rm for \;\; all} \;\; n \geq 1
 \ee
and for $\mu_q > m_q^{(T)}$, we find
 \be
\label{susp}
 {\p^2 P \ov \p \mu_q^2} \propto \le({\p \mu_q \ov \p \ep}\ri)^{-1} = {1 \ov A - B - B \log \ep}
 \ee
and
\be
 {\p^3 P \ov \p \mu_q^3} \propto - {\p^2 \mu_q \ov \p \ep^2} \le({\p \mu_q \ov \p \ep}\ri)^{-2} ={B \ov \ep} {1 \ov (A - B - B \log \ep)^2}
 \ee
In the limit $\ep \to 0$, $ {\p^2 P \ov \p \mu_q^2}  \to 0$ and
 ${\p^3 P \ov \p \mu_q^3} \to \infty$. This implies that there is
a third order phase transition since the third derivative
of the pressure is discontinuous across the phase boundary.
Also since ${\p P^2 \ov \p \mu_q^2} > 0$,
the system is thermodynamically stable.

Note that exactly for $B(T) = 0$ the
transition becomes second order, since (\ref{susp}) is then
discontinuous across the phase boundary. This happens
at two places, the first is the zero temperature critical point
which was studied in~\cite{Karch:2007br}. The second
is a new critical point at $(T,\mu)=(T_c,\mu_c)$ where
$T_c$ is given in (\ref{eq:Tc}) and $\mu_c = m_q^{(T=T_c)} \approx .0418 m_q$.
We will discuss the phase structure around this point
in the next subsection.


In Fig.~(\ref{fig:phase}) we display
the phase diagram in the $\mu_q-T$ plane, where this 3rd order
phase transition is the predominant feature.

\subsection{First order phase transition for $T>T_c$} \label{sec:4b}

Beyond the critical point $T>T_c$ we find that $B(T)$ becomes negative.
From~(\ref{susp}) this implies
that ${\p^2 P \ov \p \mu_q^2}$ becomes negative in the limit
of small $\ep$ indicating a thermodynamic instability.

To see what happens beyond $T_c$, let us plot $\mu_q$ at fixed
$T$
as a function of $\ep$. As indicated in the left plot of Fig.~\ref{fig:gmu}
the curve drops below $\mu_q = m_q^{(T)}$. Since we expect
the curve will go up again for sufficiently large density, there
must be at least one minimum somewhere, which we
call will $\mu_{\rm min}$. This minimum satisfies $\mu_{\rm min} <
m_q^{(T)}$. That is there will be multiple black hole embeddings
for a fixed $\mu_q > \mu_{\rm min}$ and which appear
as a function of $\mu_q$
before one reaches $\mu_q = m_q^{(T)}$. This allows
for the possibility of a first order phase transition
from the Minkowski embedding with $\ep=0$ to a black
hole embedding with $\ep=\ep_d(T)\neq0$  for some
chemical potential $\mu_q = \mu_d(T)$ 
between $\mu_{\rm min} < \mu_d < m_q^{(T)}$.
In this way one circumvents the thermodynamically unstable solution.

\begin{figure}[h!]
\centerline{\hbox{\psfig{figure=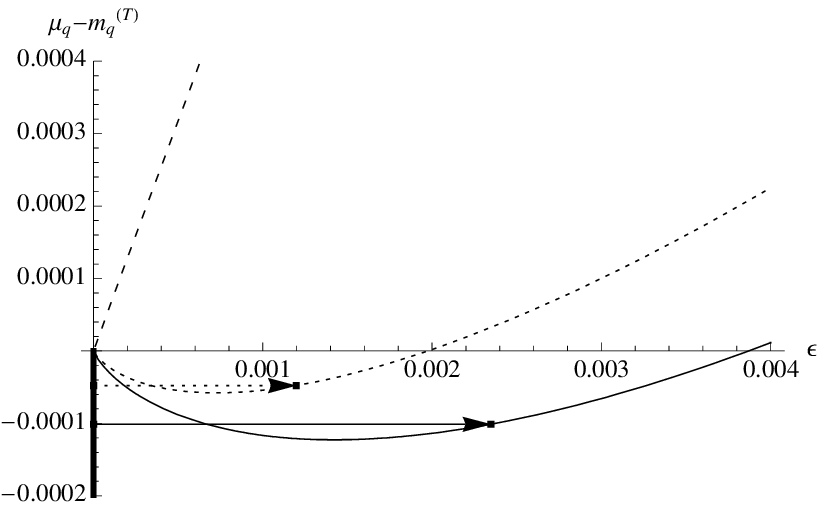,width=9cm}
\psfig{figure=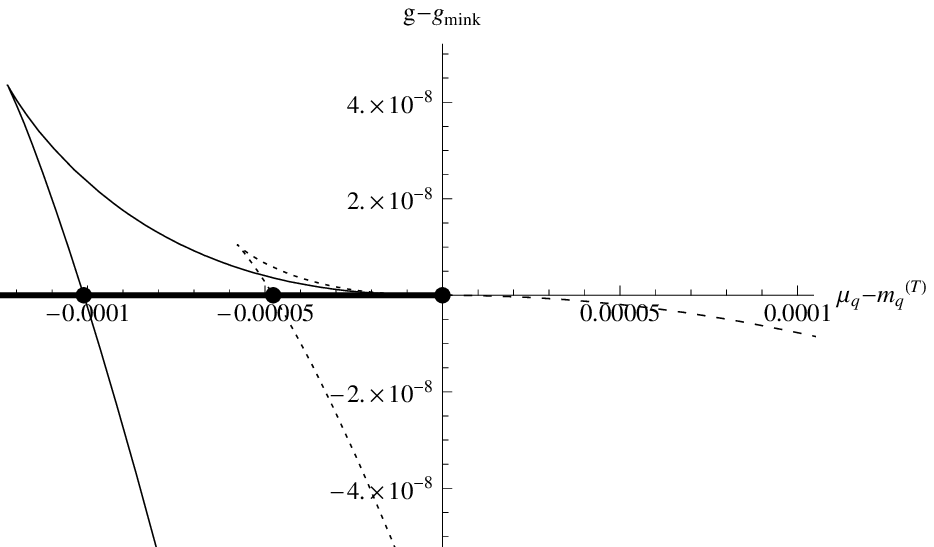,width=9cm}
}}
\caption{(\emph{left}) The quark chemical potential as a function of quark density,
for three temperatures $T/T_d= .980, .995, .996$. The first
(long dashes) is below $T_c$ and has no minimum. For the
other two we have indicated the chemical potential at
which the transition occurs.
\emph{(right)} The grand potential
density (related to the pressure by $P=-g$) as a function of chemical potential
for the same temperatures as the left figure.
\label{fig:gmu}}
\end{figure}

Although this discussion was quite general for $T>T_c$, we can
make it more explicit by considering temperatures $|T-T_c| \ll T_c$,
then the minimum and the other black hole embedding still lie at small enough values of $\ep$ that~\eqref{muF} applies with only small
corrections. The left plot of~Fig.~\ref{fig:gmu} gives the behavior of the chemical potential as a function of quark density near $T_c$ and the right plot of Fig.~(\ref{fig:gmu}) demonstrates
the swallow tail behavior of the pressure for
this first order transition.  
Using~\eqref{muF}, the first order transition point $\mu_d$ and the 
discontinuity in the density $\ep_d$ can be expressed in terms of 
the functions $A(T)$ and $B(T)$ as
\begin{equation} \label{roto}
\ep_d(T) \approx \exp\left(\frac{A-B/2}{B} \right) \qquad
\mu_d(T) - m_q^{(T)} \approx \ha B \ep_d(T)
\end{equation}
and very close to $T_c$ we can expand $A(T)$ and $B(T)$ 
\begin{equation}
A = A_c + \mathcal{O}(T - T_c), \qquad
B = - B'_c (T - T_c) + \mathcal{O}(T-T_c)^2
\end{equation}
to find 
\begin{equation}
\eps_d(T) \approx \exp\left( - \frac{A_c}{B'_c (T - T_c)} \right)
= \exp( - .12 T_d /(T-T_c) )
\end{equation}
In Fig.~(\ref{fig:Teps}) we plot $\ep_d$ and $\mu_d$ as functions of $T$ near $T_c$ using~\eqref{roto}.

\begin{figure}[h!]
\centerline{
\psfig{figure=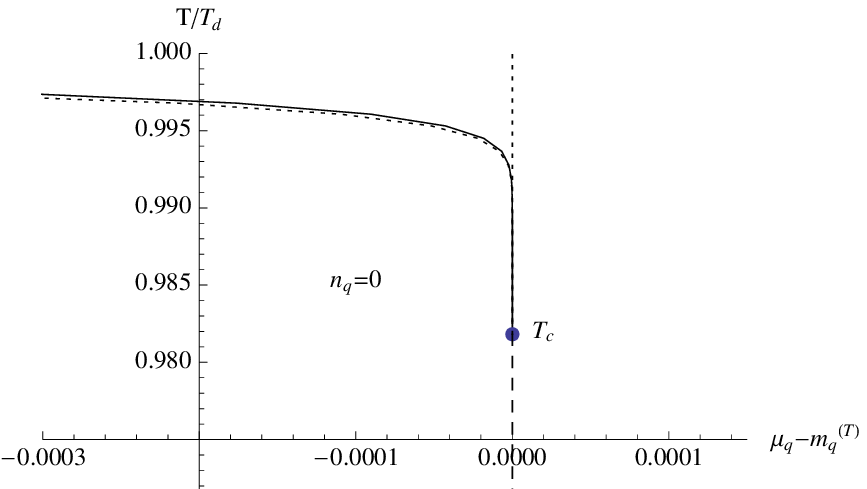,width=9cm}
\psfig{figure=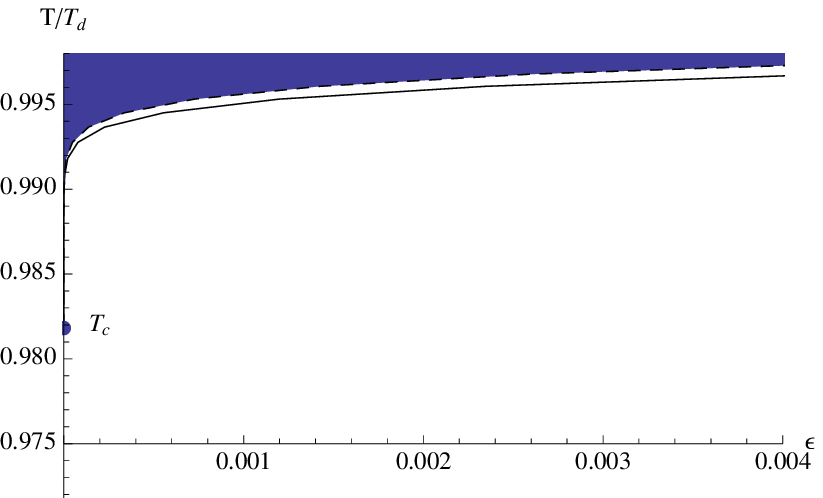,width=9cm}
}
\caption{(\emph{left}) The first order phase transition line (solid) close
to the critical point, in the $(\mu_q-m_q^{(T)})-T$ plane.
Also shown (dashed) is the region where multiple
embeddings are available at fixed $\mu_q$. (\emph{right}) The behavior of
$\eps_{d}(T)$ above the critical temperature
The shaded region represents
the onset of a thermodynamic instability. In the
canonical ensemble this region is
circumvent as usual by the Maxwell construction.
\label{fig:Teps}}
\end{figure}

To conclude this section we note that our analysis near $T_c$ 
relies on the small density behavior~\eqref{muF}. An implicit assumption 
in our discussion is that various curves in the left plot of Fig.~\ref{fig:gmu} does not 
turn back down at larger densities. 
While our small $\ep$ analysis itself does not rule out this possibility, we believe it does not happen for temperature sufficiently close to $T_c$ from our own numerical study of embedding solutions and from the finite density analysis of~\cite{Kobayashi:2006sb}.\footnote{At some higher temperature, this could happen.} Our analysis of the first order phase transition cannot be trusted for temperatures too high above $T_c$ - since then the density
at the first order phase transition will lie out of the range of applicability of~\eqref{muF}. 
Note that~\eqref{muF} is likely to break down much earlier than simply when $\ep \sim 1$.
The reason for this is that at some temperature multiple
embeddings for a fixed \emph{density} should appear,
something which our analysis has not allowed for. This
is clearly the case (as discussed earlier) for zero density,
where there exists a temperature  $T_{bh} < T_d$ above which multiple embeddings are allowed
(the original Minkowski embedding and two new black hole embeddings).
\footnote{Note that $T_{bh} = 0.9975 T_d$ is far enough away
from $T_c = .982 T_d$ that one can trust that multiple embeddings do
not appear then. } One
expects this to be the case for small densities.
Indeed these new embeddings will be continuous deformations of black hole embeddings
away from zero density, so our starting point (Minkowski embeddings)
is not good for seeing these multiple embeddings.
As we will discuss later the reason for this break down is
closely related to how this line of first order phase
transition connects to the zero density dissociation
transition.

%

\section{Conclusions and Discussions} \label{sec:con}

In this paper we showed that in the plane of temperature v.s. baryon chemical potential there is a critical line of third order phase transition which ends at a tricritical point after which the transition becomes first order. The critical behavior at the critical line is given by~\eqref{muF} which contains an intriguing logarithmic behavior. It would be interesting to have a microscopic understanding of this behavior, and more generally the structure of the whole phase diagram. In particular, it would be interesting to see whether the logarithmic behavior is related to the spiral behavior observed in~\cite{Mateos:2006nu,Mateos:2007vn} at zero chemical potential. 
Below we discuss in more detail two open issues of our investigation. 

\subsection{Connection to the  dissociation transition}

To complete the phase diagram, we need to address whether 
the first order phase transition slightly above $T_c$ discussed in sec.~\ref{sec:4b} connects to the first order phase transition along the vertical axis at $T=T_d$ 
and  $\mu_q=0$, and if yes, how. Note that such a conclusion
is suggested by numerical work in \cite{Ghoroku:2007re}
and is consistent with the numerical work in \cite{Mateos:2007vc}.

For convenience of discussion let us first briefly review 
some important aspects of the transition at $\mu_q =0$. At a low temperature 
there is only Minkowski embedding. As one increases $T$ to a temperature $T_{bh}$, 
two new black hole embeddings appear and when one further increases the temperature 
to $T_d > T_{bh} = 0.997 T_d$, a first 
order phase transition occurs, above which one of the black hole embedding thermodynamically dominates over the Minkowski embedding. For all the embedding solutions, the baryon density is zero. Thus the discontinuity $\ep_d (T)$ at the 
phase transition is trivially zero.

We now return to the question at hand,
which 
unfortunately our analysis for small density 
cannot directly answer. 
To see this if we were to extrapolate our 
analysis of the 1st order phase transition line 
(the solid line in the left figure of Fig.~\ref{fig:Teps}), 
we find that the line $\mu_d(T)$ crosses the $\mu_q=0$ axis at some 
temperature. This temperature is slightly higher than $T_d$
and hence would seem to give an estimate of $T_d$ itself,
subject to $\ep$ corrections.
This small difference from $T_d$ 
might not seem like a problem, and one might guess we have
a nice description of the physics of the zero density dissociation
transition. However it is qualitatively
wrong, since we also find $\epsilon_d(T)$ grows monotonically along this 
line. This is qualitatively wrong because on the axis $\mu_q =0$, 
the phase transition should occur at zero density. 
This demonstrates that~\eqref{muF} must have broken down quite 
drastically somewhere before
$\mu_d(T) =0$ is reached. 

\begin{figure}[h!]
\centerline{
\psfig{figure=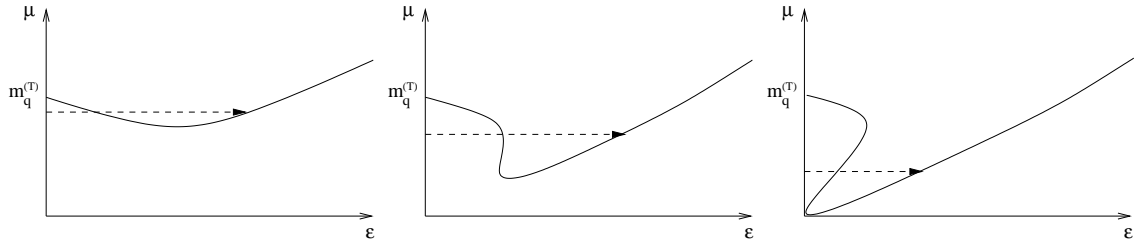,width=15cm}
}
\caption{The qualitative behavior of $\mu_q (\ep)$ at different temperatures in order for the 1st order phase transition near $T_c$ discussed in last section to connect to the transition at $\mu_q=0, \eps =0 $. 
Left plot: for a temperature $T$ slightly above $T_c$. 
Middle plot: for a temperature $T$ above $T_c$ and below $T_{bh}$. Right plot: for $T \geq T_{bh}$. 
\label{fig:trend}}
\end{figure}
 
We will now consider the simplest possibility, i.e. 
\emph{assuming} that 1st order transition line from $T_c$ does smoothly connect
with the transition at zero $\mu_q$. For this to happen  the transition density $\epsilon_d(T)$ should first increase as we increase $T$ from $T_c$ 
(consistent with the behavior derived
from~\eqref{muF})
and then decrease to zero as $T_d$ is approached. In turn this implies that the minimum of the curve $\mu_q (\ep)$ should approach the $\ep =0$ axis as the temperature is increased to $T= T_{bh}$. In~Fig.\ref{fig:trend} we plot qualitatively, at different temperatures, what is required of the curve 
$\mu_q (\ep)$ for this to happen. In particular, the minimum of the curve will hit the origin $\ep = \mu_q =0$, exactly at the temperature $T=T_{bh}$,
where two black hole embeddings appear at zero density. For temperatures above $T>T_{bh}$ moving towards $T=T_d$  the curve
$\mu_q (\ep)$  is similar to that at $T=T_{bh}$. 
However now the point at which the 1st order phase transition
occurs (which we have called $(\mu_d, \eps_d)$)
should move towards the origin.
Note that plots of form of Fig.~\ref{fig:trend}
were found numerically in~\cite{Nakamura:2007nx}.


An important feature of the last two plots~Fig.\ref{fig:trend} is that 
when temperature is sufficiently high, $\mu$ can be multi-valued for a fixed $\ep$. For $T > T_{bh}$, this is consistent with our expectation that there should be three black hole embeddings at small $\ep$, two from small density perturbations of the black hole embeddings at zero density one from the small perturbation of the Minkowski embedding which we discussed earlier in this paper. We have done some preliminary study of the behavior 
of the $\mu-\eps$ plot near the origin for $T \sim T_{bh}$ from perturbing  
two black hole embeddings at zero density and confirmed the qualitative behavior presented in the last two plots. But the extrapolation to the branch we studied in this paper requires studying finite density solution and that part of the curve 
is pure speculation at the moment. We leave the full exploration to future work.

Since, for high enough temperatures,
there are multiple embeddings with
different chemical potentials and the same density, if we
examine the thermodynamics
in the canonical ensemble,
there will be an interesting
set of phase transitions unrelated to the one identified above
that can occur between embeddings
with \emph{different} chemical potentials, but the same
density. These phase transitions are quite strange, and
as discussed in \cite{Kobayashi:2006sb} occur between solutions
which are potential thermodynamically unstable.
These transitions should involve embeddings with
chemical potentials higher than that at which the 1st order phase
transition in the grand canonical ensemble occurs. Hence
such transitions 
will be hidden from the point of view of the grand canonical
ensemble.
Since thermodynamics should be consistent in the
two different ensembles one would expect that a proper
consideration of the canonical ensemble using
for example, the Maxwell construction of inhomogeneous
phases, should
remove these differences.

\subsection{Transition at finite $\lam$}

\begin{figure}[h!]
\centerline{
\psfig{figure=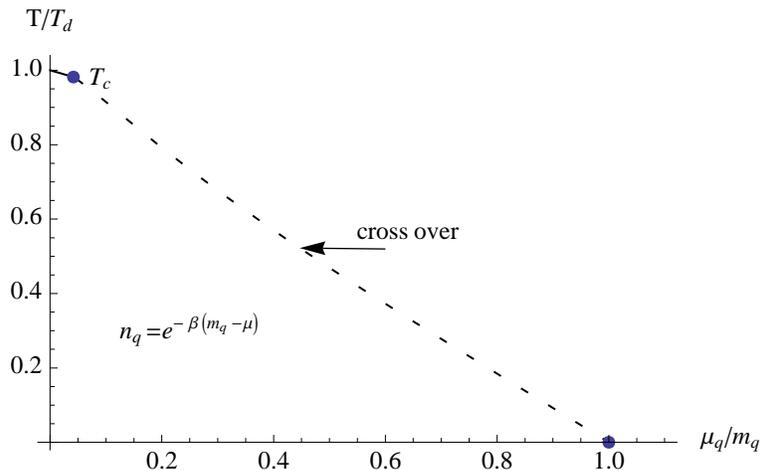,width=10cm}
}
\caption{Possible phase diagram at finite $\lam$. The lower
region consists of a Boltzmann gas of quarks and anti-quarks. The third
order phase transition of Fig.~\ref{fig:phase} is potentially smoothed to a cross
over.
\label{fig:cross}}
\end{figure}

Another important question is what could happen to the phase diagram at finite 't Hooft coupling $\lam$.  At finite $\lam$, as pointed out in~\cite{Faulkner:2008qk} and commented upon earlier in~\cite{Mateos:2007vc}, for any $\mu_q \neq 0$ at any finite temperature, the baryon
charge density is in fact \emph{nonzero}, given by a dilute Boltzmann gas of quarks.
Thus in the phase diagram Fig.~\ref{fig:phase}, the region below the critical line also has a finite density. This is of course
expected for a deconfined plasma. In particular the value of quark density at the transition line should be nonzero and continuous at finite $\lam$. This immediately raises the possibility whether the transition is smoothed into a crossover at any finite $\lam$. To have a definite answer to this question requires
summing over the instanton contributions in~\eqref{inst} for all $n$ and then taking the limit $\mu_q \to m_q^{(T)}$ in the resulting expression. This is certainly beyond the scope of the current paper. In Fig~\ref{fig:cross} we plot qualitatively the structure of the phase diagram if the transition is smoothed out.

Here we mention another indirect indication that the transition might be smoothed out.
In~\cite{Faulkner:2008qk} we have argued that
at any finite $\lambda$, in the region below the transition line in the phase diagram, the mesons have a width proportional to the sum of quark and anti-quark densities,
\begin{equation}
\label{width}
 \Ga = {32 \pi^3 \sqrt{\lam} \ov N_c  m_q^2} |\psi (\th =0)|^2 \, (n_+ + n_-) \ .
\end{equation}
where $n_{\pm} \propto \exp( - (m_q^{(T)}\pm \mu_q)/T)$ and
are exponentially small in $\sqrt{\lambda}$ as argued in sec.\ref{sec:ins}.
$\psi(\th=0)$ is the meson wave function at the tip of the brane. 
We have also performed~\cite{faulk} a calculation of the meson width on the critical line 
from above and found at small densities exactly the same answer as~\eqref{width}.\footnote{The meson widths at small density have also been studied numerically in~\cite{Myers:2008cj}.}  
Note that one of the key signatures of the phase transition
at zero density (in fact potentially the defining signature)
is the spectrum of mesons (quark and anti-quark bound states)\cite{Hoyos:2006gb,Mateos:2007vn}.
Here we find the meson widths and thus the spectrum are continuous across the critical line
indicating that the transition might be smoothed out.

\begin{acknowledgments}

We thank D.~Mateos, R.~Myers, K.~Rajagopal, 
V.~Kumar and T.~Senthil for useful discussions.
Research supported in part by
the DOE
under
contracts
\#DF-FC02-94ER40818.
HL is also supported
in part by the A.~P.~Sloan Foundation and the 
DOE OJI program.  

\end{acknowledgments}

\appendix

\section{The zero temperature solution at small densities} \label{app:zerot}

One way to motivate the scaling of the inner region $\rho = \sigma \ep^{1/2}$
is to look at the zero temperature solution in detail, where the exact
solution is known.

This solution is:
\begin{equation}
\label{eq:zerosoln}
y'(\rho) = \frac{c}{\sqrt{ \rho^6 + \ep^2 - c^2}}
\qquad a'(\rho) = \frac{\ep}{c} y'(\rho)
\end{equation}
where $c$ is an integration constant related to the
expectation of the mass operator dual to the field $y$
on the gravity side. The boundary conditions are $y(0) = a(0) = 0$.

We can fix $c$ in terms of the quark mass, or equivalently in terms
of $L$. After rescaling as (\ref{scel}) by $L_0 = L$ the condition
$y(\infty) = 1$ fixes the quark mass.
Integrating (\ref{eq:zerosoln}) one finds for $c$,
$$
\kappa^3 c^3 = ( \ep^2 - c^2 )
$$
where $\kappa$ is a number defined by the integral,
$
\kappa = \int_0^\infty dx/(x^6 + 1)$.
For small densities the condensate $c$ approaches $\ep$ as,
\begin{equation}
c \approx \ep - (\kappa/L)^3 \ep^2 \,, \qquad
\ep^2 - c^2 \approx \ep^3 ( \kappa/L)^3
\end{equation}
Note that by small densities we really mean
$\epsilon \ll L^3$ or
equivalently $n_q/(N_c \sqrt{\lambda}) \ll m_q^3$.

From (\ref{eq:zerosoln}) it is clear
the cross over scale
for $y(\rho)$ between a flat Minkowski embedding
and the curved solution going to the (AdS) horizon is $\rho^6 \sim
\ep^2 -c^2$ or $\rho \sim \ep^{1/2}$ at small densities.
This is the scaling that we set out to demonstrate.

Rescaling $\sigma = \rho/\ep^{1/2}$
we find the zero temperature \emph{inner} solution becomes,
\begin{equation}
\label{eq:zerotinner}
\frac{d Y_0}{d \sigma} \approx \frac{1}{ \sqrt{\sigma^6 +
(\kappa)^3 } }
\end{equation}
In this limit the chemical potential to order $\epsilon$ receives contributions
only from inner region with the result
\begin{equation}
\label{eq:zt}
\mu_q = \frac{L}{2 \pi \alpha'} \left( 1  +  \frac{1}{2} \kappa^3 \ep + \mathcal{O}(\ep^2) \right)
\end{equation}

\section{Detailed analysis of the chemical potential}\label{app:C}

In this appendix we give some details in the small $\ep$-expansion of $\mu_q$, equation~\eqref{mu}, which we copy here for convenience
  \be \label{mun}
\mu_q = {L_0 \ov 2 \pi \apr} \ep\int_{\rho_c}^\infty d \rho \, {\sqrt{fq (1+y'^2)} \ov \sqrt{\ep^2 + \rho^6 q^3}} \ .
 \ee
Various functions in~\eqref{mun} were introduced in~\eqref{rfdr}. We split the above integral into those over the inner and outer regions
 \be
 \mu_q = \int_{\rho_c}^{\rho_\Lam} (\cdots) + \int_{\rho_
 \Lam}^{\infty} (\cdots) \equiv \mu_{\sI} + \mu_{\sO} \
 \ee
which can then be expanded in terms of $\ep$ using the expansions~\eqref{in},~\eqref{out} of $y(\rho)$ in inner and outer regions respectively.

Let us first look at $\mu_{\sO}$. We find that it starts with order $\ep$
\be
  \mu_\sO = \ep \, \mu_{\sO}^{(1)} + O(\ep^2)
  \ee
with
 \be
 { 2 \pi \apr \ov L_0 } \mu_\mathcal{O}^{(1)}
=    \int_{\rho_\Lambda}^\infty
d\rho \frac{\left( 1 - \eta^4/u^4\right)\sqrt{ 1 + y_0'^2}}
{\left(1 + \eta^4/u^4\right)^2 \rho^3 }
\ee
and $u^2 = \rho^2 + y_0^2 (\rho)$. Using the expansion~\eqref{outer} of $y_0$ at
small $\rho$, we find that the integral contains a quadratic and a logarithmic divergent term in the limit $\rho_\Lam \to 0$, given by
  \be \label{aoo1}
 { 2 \pi \apr \ov L_0 } \mu_{\sO}^{(1)} = \frac{ 1- \eta^4}{2 ( 1 + \eta^4)^2 \rho_\Lambda^2 } - \frac{ 2  \eta^4 (3 - \eta^4 + 3 \eta^8)}{(1 - \eta^4)(1 +\eta^4)^4}
\log(\rho_\Lambda)  + K_\sO
 \ee
where $K_\sO$ denotes the finite part of the integral. It can be defined more precisely by
 \be \label{ond}
K_\sO  = \lim_{\rho_\Lambda \rightarrow 0 }  \left[\int_{\rho_\Lambda}^\infty
d\rho \frac{\left( 1 - \eta^4/u^4\right)\sqrt{ 1 + y_0'^2}}
{\left(1 + \eta^4/u^4\right)^2 \rho^3 }
- \frac{(1- \eta^4)}{2 ( 1 + \eta^4)^2 \rho_\Lambda^2 }
- \frac{ 2 \eta^4 (3 - \eta^4 + 3 \eta^8)}{(-1 + \eta^4)(1 +\eta^4)^4}
\log(\rho_\Lambda) \right]
 \ee
Once $y_0 (\rho)$ is known numerically, $K_\sO$ can be calculated numerically.

We now look at $ \mu_{\sI}$ which can be expanded as
\be
 \mu_{\sI} = \mu_{\sI}^{(0)} + \ep \mu_{\sI}^{(1)} + \cdots
 \ee
The first term in the above has been discussed explicitly in the main text. Here we give some details for computing $\mu_{\sI}^{(1)}$, which can be written as (derivatives are w.r.t. $\sig$)
\be \label{ine1}
  { 2 \pi \apr \ov L_0} \mu_{\sI}^{(1)}  = \int_{\sig_0}^\Lam d \sig \, \le(M  + \p_\sig \le(Y_1  \frac{\left(Y_0^4 - \eta^4\right)}
 {Y_0^2 \left( \eta^4 + Y_0^4 \right)^{1/2}}  \ri) \ri),
 \ee
where
 \be
M = \frac{ Y_0^4 - \eta^4}{2 Y_0' Y_0^2 \sqrt{\eta^4 + Y_0^4}}
 + \frac{\sig^2 \eta^4(\eta^4 + 3 Y_0^4) Y_0'}{ Y_0^4\left( \eta^4
 + Y_0^4 \right)^{3/2}} + \frac{ \sig^6\left(\eta^4 - Y_0^4 \right)
 \left( \eta^4 + Y_0^4 \right)^{5/2} Y_0'}{2 Y_0^{14}}
\ee
The total derivative term in~\eqref{ine1} is nonzero only at the upper end and gives
 \be \label{tot}
 {\rm{total}} =
 \frac{\eta^8}{(1+ \eta^4)^{3/2}} \Lambda^2
 + \frac{(1-\eta^4) a_1}{\sqrt{1+ \eta^4}}
 + \frac{ \eta^{12} ( 3+ \eta^4) - 14 \eta^8 \log(\Lambda) }{(-1 + \eta^4)(1+\eta^4)^4}
 \ee
where we have used the expansions~\eqref{y0E} and~\eqref{eq:inassym1}. The integration over $M$ in~\eqref{ine1} is also divergent in the limit $\Lam \to \infty$. The divergent terms can be extracted using the expansion~\eqref{y0E} and one finds that
 \be \label{tot1}
\int_{\sig_0}^\Lam d \sig \, M =
 - \frac{\eta^8}{(1+ \eta^4)^{3/2}} \Lambda^2
- \frac{2 \eta^4 ( 3 - 8 \eta^4 + 3 \eta^8)}{(-1+ \eta^4) (1+ \eta^4)^4}
\log(\Lambda) + M_{\rm f}
\ee
where $M_{\rm f}$ denotes the finite part of the integral and can be defined more precisely as
  \be \label{u2}
 M_{\rm f} = \lim_{\Lam \to \infty} \le( \int_{\sig_0}^\Lam d \sig \, M
 + \frac{\eta^8}{(1+ \eta^4)^{3/2}} \Lambda^2
+ \frac{2 \eta^4 ( 3 - 8 \eta^4 + 3 \eta^8)}{(-1+ \eta^4) (1+ \eta^4)^4}
\log(\Lambda) \ri)
\ee
Adding~\eqref{tot} and~\eqref{tot1} together we find that the quadratic divergences in $\Lam$ cancel and
 \bea \label{aoo3}
 && { 2 \pi \apr \ov L_0} \mu_\sI^{(1)} \cr
&&= \frac{2 \eta^4 ( 3 -  \eta^4 + 3 \eta^8)}{(1- \eta^4) (1+ \eta^4)^4}
\log(\Lambda) +  {7 \eta^8 \log \ep \ov (1 - \eta^4)(1+\eta^4)^4} +  \frac{1 - \eta^4}{
\sqrt{1 + \eta^4}} b_1 - \frac{ \eta^{12} ( 3 + \eta^4)}{(1 - \eta^4)(1+ \eta^4)^4}  +
M_{\rm f} \nonumber \\
 \label{rr1}
\eea
where we have used~\eqref{inC1} and $b_1$ was introduced in~\eqref{ex1}. Equation~\eqref{rr1} gives~\eqref{oo1} with the definition
 \be
K_\sI  \equiv  \frac{1 - \eta^4}{
\sqrt{1 + \eta^4}} b_1 - \frac{ \eta^{12} ( 3 + \eta^4)}{(1 - \eta^4)(1+ \eta^4)^4}  +
M_{\rm f}
\ee
$K_\sI$ depends on the full solution and can be computed numerically.

\end{document}